%% file: letter.tex
\documentclass[aps,prl,showpacs,notitlepage,nofootinbib,superscriptaddress,floatfix,showkeys,twocolumn]{revtex4-2}

\usepackage{blindtext}
\usepackage{hyperref}
\usepackage{amsmath,amssymb}
\usepackage{float}
\usepackage{microtype}
\usepackage{graphicx}
\usepackage{bm}
\usepackage{latexsym}
\usepackage{epsfig}
\usepackage{psfrag}
\usepackage{color}
\usepackage[dvipsnames]{xcolor}
\usepackage{subfigure}
\usepackage[section]{placeins}

\def\nn{\nonumber} 
\def\f{\frac}

\def\l{\left}
\def\r{\right}
\def\d{{\mathrm{d}}}

\def\Mpl{M_{_{\mathrm{Pl}}}}
\def\HI{H_{_\mathrm{I}}}
\def\ei{\eta_{\mathrm{i}}}
\def\ee{\eta_{\mathrm{e}}}
\def\e1i{\epsilon_{1\mathrm{i}}}
\def\ps{\mathcal{P}_{_{\mathrm{S}}}}
\def\bs{\mathcal{B}_{_{\mathrm{S}}}}
\def\ns{n_{_{\mathrm{S}}}}
\def\mpcinv{\mathrm{Mpc}^{-1}}
\def\kdip{k_{\mathrm{dip}}}

\def\Ts{T_{_{\mathrm{S}}}}
\def\Tcmb{T_{_{\mathrm{CMB}}}}
\def\Tk{T_{_{\mathrm{K}}}}
\def\yc{y_{\mathrm{c}}}

\allowdisplaybreaks[1]

\def\vk{\bm k}
\def\vka{{\bm k}_{1}}
\def\vkb{{\bm k}_{2}}
\def\vkc{{\bm k}_{3}}
\def\ei{\eta_{\rm i}}
\def\ee{\eta_{\rm e}}

\def\cR{{\mathcal R}}

\begin{document}

\title{Observing nulling of primordial correlations via the $21\, \mathrm{cm}$ 
signal}
\author{Shyam~Balaji}
\email{sbalaji@lpthe.jussieu.fr}
\affiliation{Laboratoire de Physique Th\'{e}orique et Hautes Energies (LPTHE),\\
UMR 7589 CNRS \& Sorbonne Universit\'{e}, 4 Place Jussieu, F-75252, Paris, France}
\author{H.~V.~Ragavendra}
\email{ragavendra.pdf@iiserkol.ac.in}
\affiliation{Department of Physical Sciences, Indian Institute of 
Science Education and Research Kolkata, Mohanpur, Nadia~741246, India}
\author{Shiv K. Sethi}
\email{sethi@rri.res.in}
\affiliation{Raman Research Institute, C.~V.~Raman Avenue, Sadashivanagar, 
Bengaluru~560080, India}
\affiliation{Centre for Strings, Gravitation and Cosmology, Department of 
Physics, Indian Institute of Technology Madras, Chennai~600036, India}
\author{Joseph~Silk}
\email{silk@iap.fr}
\affiliation{Institut d'Astrophysique de Paris, UMR 7095 CNRS \& Sorbonne 
Universit\'{e}, 98 bis boulevard Arago, F-75014 Paris, France}
\author{L.~Sriramkumar}
\email{sriram@physics.iitm.ac.in}
\affiliation{Centre for Strings, Gravitation and Cosmology, Department of 
Physics, Indian Institute of Technology Madras, Chennai~600036, India}

\begin{abstract}
The $21\,\mathrm{cm}$ line emitted by neutral hydrogen (HI) during the Dark 
Ages carries imprints of pristine primordial correlations.
In models of inflation driven by a single, canonical scalar field,  we show 
that a phase of ultra-slow-roll can lead to a {\it null}\/ in {\it all}\/ the 
primordial correlations at a specific wavenumber~$\kdip$.
We consider scenarios wherein the null in the correlations occurs over wave 
numbers $1 \lesssim \kdip \lesssim 10\,\mathrm{Mpc}^{-1}$, and examine the 
prospects of detecting such a damping in the HI signal due to the nulls at 
the level of power and bi-spectra in future observational missions.
\end{abstract}
\maketitle


\noindent
\underline{\it Primordial correlations and $21\, \mathrm{cm}$ observations:}\/~Cosmic 
inflation remains the most attractive paradigm for the generation of 
primordial perturbations.
On large scales, e.g. over $10^{-5}\lesssim k \lesssim 1\,\mathrm{Mpc}^{-1}$, 
the primordial scalar power spectrum as generated in some of the popular models 
of slow roll (SR) inflation is remarkably consistent with the cosmic microwave 
background (CMB) anisotropies  and large-scale structure (for a comprehensive 
list of inflationary models consistent with the Planck data, see
Refs.~\cite{Martin:2013tda,Martin:2013nzq}).
However, on smaller scales, e.g. $k\gtrsim 1\, \mathrm{Mpc}^{-1}$, the constraints
on the primordial scalar power spectrum are considerably weaker.
Since the discovery of gravitational waves from merging binary black holes, the 
weaker constraints over small scales have been exploited to examine inflationary
models which enhance power on these scales. This  leads to   significant production of 
primordial black holes and generation of secondary gravitational waves of 
observable strengths~\cite{Latif:2016qau,Garcia-Bellido:2017mdw,Ballesteros:2017fsr,
Germani:2017bcs,Kannike:2017bxn,Carr:2018rid,Dalianis:2018frf,Ragavendra:2020sop,
Ragavendra:2020vud,Gow:2020bzo,Garcia-Bellido:2016dkw,Ragavendra:2021qdu,
Balaji:2022rsy,Balaji:2022dbi}.
\enlargethispage*{0.75cm}

Often, in single field models of inflation involving the canonical scalar field, 
a phase of ultra-slow-roll (USR) is invoked to enhance the scalar power on small 
scales. The first SR parameter $\epsilon_1$ exponentially decreases during such 
a phase,  resulting in large values for the second and higher order SR
parameters~\cite{Tsamis:2003px,Kinney:2005vj,Garcia-Bellido:2017mdw,
Ballesteros:2017fsr,Germani:2017bcs}.
Such a departure from SR inflation leads to a peak in the inflationary 
scalar power spectrum and, generically, one finds that the power spectrum 
rises as~$k^4$ as it approaches the peak~\cite{Byrnes:2018txb,Tasinato:2020vdk,
Cole:2022xqc}.
Interestingly, just before the power spectrum rises towards the peak, a sharp 
drop in power occurs~\cite{Ozsoy:2021pws} and, if the period of USR inflation
is sufficiently long, the scalar power spectrum actually vanishes at a particular
wavenumber, which we denote as~$\kdip$~\cite{Goswami:2010qu}.
This occurs because of the fact that the mode function describing the curvature
perturbation corresponding to the wavenumber~$\kdip$ goes to zero at late times
towards the end of inflation.
It can immediately be shown that all of the higher correlations involving 
the curvature perturbation will also necessarily vanish at~$\kdip$.
In cases wherein the duration of USR is not sufficiently long, although a null 
does not arise, a sharp dip in the scalar power spectrum as well as in the 
higher order correlation functions is still encountered. 

Over the scales $1 \lesssim k \lesssim 10\,\mpcinv$, the $21\,\mathrm{cm}$
signal of neutral hydrogen (HI) from the Dark Ages carries the signatures 
of the primordial spectrum (see, e.g., Refs.~\cite{2010ARA&A..48..127M,21cm_21cen}).
In contrast to the angular spectra of the CMB, which are a convolution of the 
primordial spectra and the transfer function of the photons, the features in 
the primordial power and bi-spectra leave direct and distinct imprints in the 
HI signal. 
Therefore, an inflationary feature such as a null or a sharp dip in the primordial 
correlations may potentially be observed in HI, if the features occur over 
the corresponding scales~\cite{Munoz:2016owz}.
In this Letter, we consider specific scenarios involving a phase of USR inflation
and investigate the effects of a dip on the HI signal at the level of both power 
and bi-spectra.
If a drop in the scalar power spectrum is to occur over $1 \lesssim k \lesssim
10\,\mpcinv$, we find that the CMB at smaller wavenumbers and the spectral 
distortions at higher wavenumbers limit the rise in power on small scales, and 
hence the extent of the dip. 
We calculate the corresponding observable signatures on the power and bi-spectra
of the HI signal and discuss the prospects of observing them in future missions, 
such as a lunar array~\cite{Furlanetto:2019jso,Cole:2019zhu,Koopmans:2019wbn}.
We also point out challenges that can arise due to Poisson
fluctuations (PF)~\cite{Afshordi:2003zb, Mack:2008nv}.


\noindent
\underline{\it Nulls in inflationary correlations}:\/~We now demonstrate 
that nulls in the inflationary correlations (i.e. in the scalar power spectrum
as well as in higher order correlations) are expected to arise in scenarios 
involving a phase of USR inflation.
Consider a situation wherein a regime of USR inflation is sandwiched between 
two epochs of SR inflation.
Let $\eta$ denote the conformal time coordinate, and let the two transitions
between the three stages occur at the times~$\eta_1$ and~$\eta_2$.
Also, let the first SR parameter $\epsilon_1$ prior to the first
transition be a constant, say $\epsilon_{1\mathrm{i}}\lesssim 10^{-2}$, 
while, during the period of USR, it is given by
$\epsilon_1=\epsilon_{1\mathrm{i}}\, (\eta/\eta_1)^6$.
Since $\epsilon_1 \ll 1$ throughout the domains of interest, the 
Hubble parameter can be considered to be a constant, say, $\HI$, and 
hence the scale factor can be assumed to be of the de Sitter form.

Let us focus on the evolution of modes that leave the Hubble radius during
the initial SR regime.
In the first domain $\eta <\eta_1$, on super-Hubble scales, the mode 
function characterizing the curvature perturbation in Fourier space, say,
$f_k^{\mathrm{I}}$, can be expressed as 
\begin{equation}
f_k^{\mathrm{I}}(\eta)
=C_k+\f{D_k}{2}\,\eta^2.
\end{equation}
The constants $C_k$ and $D_k$ can be determined by matching the super-Hubble
solutions with the complete solution in the SR regime, and they are found to 
be $C_k = i\,\HI/(\sqrt{4\,k^3\,\epsilon_{1\mathrm{i}}}\,\Mpl)$
and $D_k = C_k\,k^2$.
During the USR phase, the modes function, say, $f_k^{\mathrm{II}}$, for modes 
that are already on super-Hubble scales, can be expressed as
\begin{equation}
f_k^{\mathrm{II}}(\eta)
= A_k + B_k\, \l(\f{1}{\eta^{3}}-\f{1}{\eta_1^{3}}\r).
\end{equation}
The quantities $A_k$ and $B_k$ can be determined by matching the mode 
functions and their time derivatives at the transition at~$\eta_1$ to 
obtain that $A_k = C_k\,[1+(k^2\,\eta_1^2/2)]$ and $B_k = -D_k\,\eta_1^5/3$.

When the phase of USR ends, because the wavenumber of interest 
is on super-Hubble scales, its amplitude will evidently freeze at its value
at the conformal time~$\eta_2$.
We should clarify that such a behavior can also be expected if, for $\eta>
\eta_2$, the parameter $\epsilon_1$ begins to grow leading to the termination 
of inflation.
Hence, the power spectrum is determined by the value of $f_k^{\mathrm{II}}$
at~$\eta_2$.
Upon setting $f_k^{\mathrm{II}}(\eta_2)$ to be zero, we can immediately
determine the wavenumber~$\kdip$ at which the amplitude of the curvature 
perturbation vanishes.
It is given by 
\begin{equation}
\kdip \!=\! -\f{1}{\eta_1}
\l\{\f{1}{3}\l[\l(\f{\eta_1}{\eta_2}\r)^{3}-1 \r] -\f{1}{2}\r\}^{-1/2}
\!\!\!\simeq\! \sqrt{3}\,k_1\,\mathrm{e}^{-3\,\Delta N/2},\label{eq:kdip}
\end{equation}
where the final expression has been arrived at by assuming that the epoch 
of USR is adequately long so that $\eta_1/\eta_2 \gg 1$, 
and we have set $k_1=-1/\eta_1$ (i.e. the wavenumber that leaves the 
Hubble radius at the onset of USR), while $\Delta N$ denotes
the duration of USR in e-folds.
The power spectra and all the higher order correlations involve the 
mode function $f_k$ evaluated towards the end of inflation. 
{\it Since the mode function $f_k$ corresponding to the wavenumber 
$\kdip$ vanishes at late times, any correlation function involving 
this mode necessarily vanishes as well.}\/
However, if the duration of USR is not long enough (in fact, when
$\mathrm{e}^{\Delta N}\lesssim 5/2$), then the mode function, rather 
than vanishing, settles down to a very small value at late times.
In such cases, a sharp dip is produced rather than a null in the correlation
functions, and the relation~\eqref{eq:kdip} predicts the location of the dip. 
Moreover, when the dominant term in the mode function $f_k$ 
vanishes, the sub-dominant terms can lead to a small non-zero value at $\kdip$,
resulting in a dip as opposed to a null.
\enlargethispage*{1.0cm}


\noindent
\underline{\it Inflationary models, power and bi-spectra:}\/~To illustrate 
the nulls or dips that are expected in the correlation functions, we shall 
consider two models of inflation driven by a single, canonical scalar field 
that permit a brief period of USR. 
These models should be treated as illustrative examples of inflationary 
scenarios generally considered to enhance power on small scales. 
We shall also briefly discuss a reconstructed scenario which easily allows 
us to achieve the desired background evolution and a power spectrum that is 
consistent with the constraints from the CMB on large scales.

The first model we shall consider is a model due to Starobinsky that 
is described by a linear potential with a sudden change in its 
slope~\cite{Starobinsky:1992ts,Arroja:2012ae,Martin:2011sn,Sreenath:2014nka}. 
It is one of the simplest models that leads to a regime of USR and a 
step-like feature in the scalar power spectrum.
The potential describing the Starobinsky model (ST) is given 
by~\cite{Starobinsky:1992ts,Arroja:2012ae,Martin:2011sn,Sreenath:2014nka}
\begin{equation}\label{eq:sm}
V(\phi) 
=\begin{cases}
V_0 + A_+\,(\phi-\phi_0) & \mathrm{for}~\phi > \phi_0,\\
V_0 + A_-\,(\phi-\phi_0) & \mathrm{for}~\phi < \phi_0,
\end{cases}
\end{equation}
where~$V_0$ sets the overall energy scale.
Evidently, $A_+$ and $A_-$ determine the slopes of the potential on 
either side of~$\phi_0$, and the slope is discontinuous at this point.
Though there are issues in achieving a natural end to inflation, we 
consider the model because of its analytical tractability that helps
in illustrating arguments related to features induced by USR (in this
regard, see the supplementary material).
In the model, the epoch of USR occurs when the field crosses~$\phi_0$ 
and the duration of this epoch is determined by the ratio of the slopes 
of the potential, i.e. $A_-/A_+$.
It can be shown that, in the model, $\kdip \simeq\sqrt{3\,(A_-/A_+)}\,
k_0$, where $k_0$ is the wavenumber that leaves the Hubble radius when
the field crosses~$\phi_0$~\cite{Starobinsky:1992ts,Martin:2011sn,
Ragavendra:2022mip}.
The parameters $V_0$ and $A_+$ are constrained by COBE normalization on 
the CMB scales (for values of the parameters, see the supplementary 
material, which includes Refs.~\cite{Martin:2014kja,Ragavendra:2020old}).
The constraints from spectral distortions over the wavenumbers $1 < k 
< 10^4\,\mpcinv$ limit the extent of enhancement in the power spectrum 
at small scales~\cite{Chluba:2019nxa}, and hence the duration of USR.
We choose the parameters $A_-$ and $\phi_0$ so that the rise in power
on smaller scales is consistent with the FIRAS constraints on 
$\mu$ distortions~\cite{Gow:2020bzo}. 
Also, these parameters are chosen such that the dip in the power spectrum 
occurs at wavenumbers $k\gtrsim 5 \, \rm Mpc^{-1}$ to evade bounds on the 
matter power spectrum from the Lyman-$\alpha$ data (see e.g.,
Refs.~\cite{mcquinn2016evolution,Afshordi:2003zb}).
For the parameters we work with, we find that $\kdip= 7.6\,\mpcinv$.
We should clarify that, the duration of USR in ST is 
determined by the ratio of $A_-/A_+$, which, in turn, determines the 
height of scalar power at its maximum. 
Since this amplitude is constrained by $\mu$ distortion, it imposes a 
lower bound on this ratio. 
Such a bound leads to an inadequate duration of USR, producing a sharp dip, 
instead of a null in the power spectrum.

The second model we shall consider is an inflationary scenario driven by the
Higgs field that is coupled non-minimally to gravitation~\cite{Ezquiaga:2017fvi,
Bezrukov:2017dyv,Drees:2019xpp,Ragavendra:2021qdu}. 
The model is known as critical-Higgs inflation (CH), and the effective
potential in this scenario contains a point of inflection, which leads 
to an epoch of USR thereby enhancing the scalar power over small scales. 
The potential describing the model can be written as
\begin{equation}
V(\phi) = V_0\,\f{\l[1+a\,\l(\mathrm{ln}\,z\r)^2\r]\,z^4}
{\l[1+c\,\l(1+b\,\mathrm{ln}\,z\r)\,z^2\r]^2},\label{eq:phi-Higgs}
\end{equation}
where $z = \phi/\phi_0$. 
As in the case of ST, we choose the parameters of the potential so that
the power spectrum is consistent with COBE normalization on large scales
and with the constraints from spectral distortions on smaller scales.
For the values of the parameters we work with (in this regard, see the
supplementary material), we find that a dip in the power spectrum 
occurs at around $\kdip =7.6\,\mpcinv$ and the power reaches its 
maximum amplitude at around $5.5\times10^9\,\mpcinv$.
The $\mu$ distortion arising due to this spectrum is found to be about
$2.0\times10^{-5}$, which is within the FIRAS bound~\cite{Chluba:2019nxa,
Gow:2020bzo}.

It is known that, in single field models of inflation, if the enhancement 
in power is to be achieved over wavenumbers that are close to the CMB scales, 
there can arise a tension between the  value of the scalar spectral index~$\ns$ 
in the model and the constraint on the parameter from the CMB data (for 
recent discussions, see Refs.~\cite{Iacconi:2021ltm,Karam:2022nym}).
In the ST and CH models, the value of $\ns$ at the pivot scale of $k_\ast=0.05\,
\mpcinv$ turns out to be $0.9995$ and $0.78$, respectively, which are well 
away from the mean value of~$0.96$ from Planck~\cite{Aghanim:2018eyx}.
One way to circumvent this challenge is to construct inflationary potentials 
using the desired behavior of $\epsilon_1(N)$ and, interestingly, it can be 
shown that these reconstructed potentials too contain a point of 
inflection~\cite{Ragavendra:2020sop}.
In other words, using methods of reconstruction, it is possible to arrive at 
potentials numerically that are consistent with the CMB data on large scales
and lead to a dip in the power spectrum over $1 \lesssim k \lesssim 10\, 
\mpcinv$ (for details, see the supplementary material).
\enlargethispage*{0.75cm}

\begin{figure}[!t]
\centering
\includegraphics[width=1\linewidth]{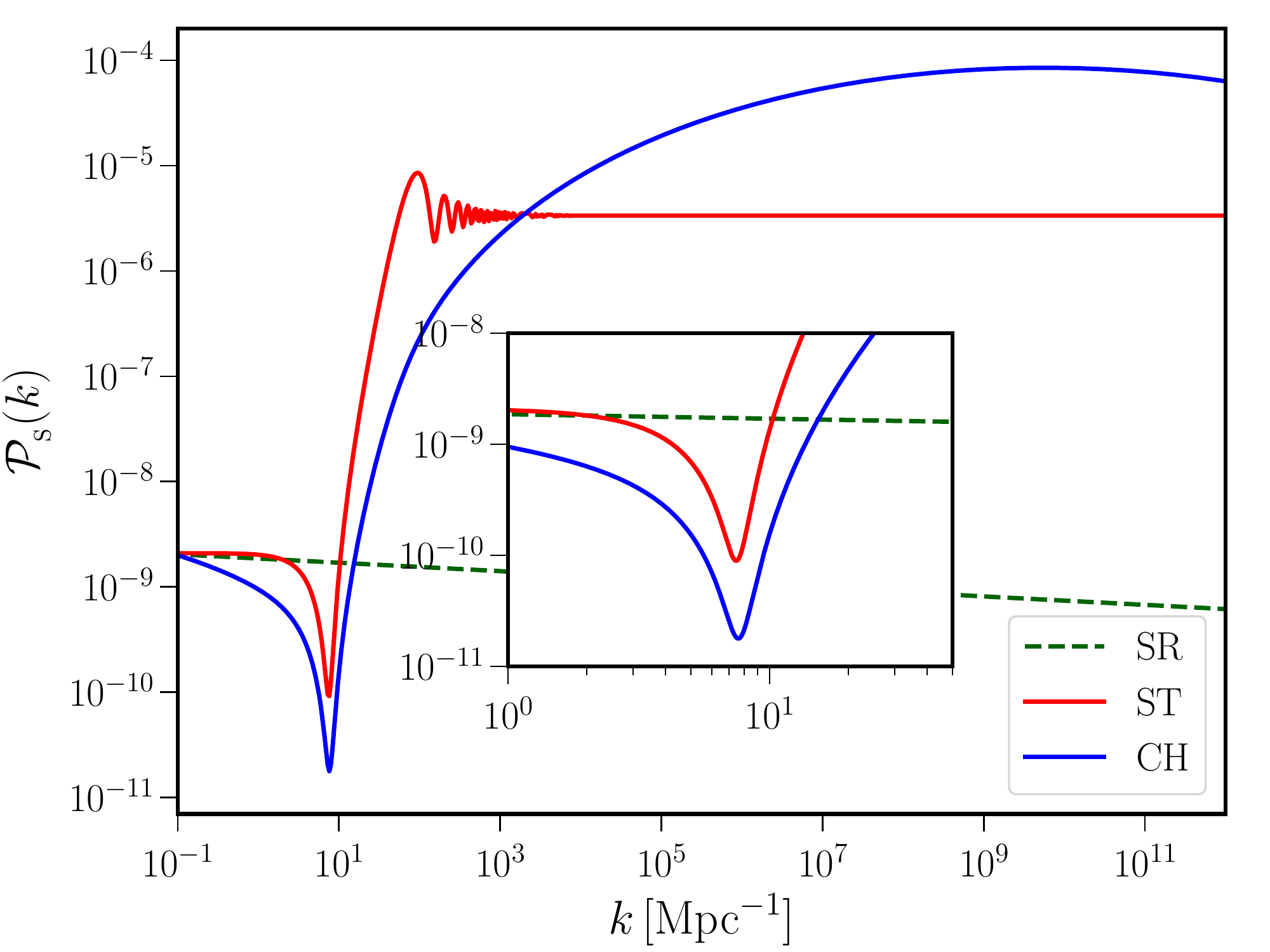}
\caption{The inflationary scalar power spectra arising in ST and CH models are 
illustrated for parameters that are consistent with the constraints on 
spectral distortions from FIRAS.
We have also plotted the nearly scale invariant spectrum that may be 
obtained from a typical SR model of inflation.
The inset highlights the dip in the spectra at $k=7.6\,\mpcinv$, and we 
work with parameters such that the dip occurs over wavenumbers where the 
HI signal is expected to be  most sensitive to the primordial power
spectrum.}
\label{fig:sps}
\end{figure}

We find that the features around the dip have the same characteristics in the 
reconstructed scenario as in the ST and CH models.
Therefore, we shall proceed by considering these models and examining their 
imprints on the $21\,\mathrm{cm}$ signal.
We evolve the background and compute the scalar power and bi-spectra numerically
(see the supplementary material for details, which includes
Refs.~\cite{Maldacena:2002vr,Seery:2005wm,Chen:2010xka,
Arroja:2011yj,Hazra:2012yn}).
In Figs.~\ref{fig:sps} and~\ref{fig:sbs} (see App.~I), we have presented the 
inflationary scalar power and bi-spectra, i.e. $\ps(k)$ and $\bs(\vka,\vkb,\vkc)$,
that arise in the ST and CH models, for the values of the parameters we have 
worked with. 

We note two related points regarding the bispectra.
Firstly, in contrast to the power spectrum which is a positive definite quantity,
the bispectra can cross zero when departures from slow roll arise.
Hence, it may vanish at locations other than~$\kdip$.
However, these nulls are dependent on the nature of the integrals 
involved in the computation of the bispectrum and may not be observed at the same
locations in the higher order correlations. 
Therefore, they are not as generic as the dip of interest, which will be located
at $\kdip$ in the higher order correlations as well. Hence,
in models wherein a deviation from slow roll arises due to an epoch of USR, 
the bispectra are {\it guaranteed}\/ to exhibit a sharp dip at~$\kdip$.
Secondly, note that the power and bispectra in the Starobinsky  model rise 
more sharply than in the Higgs model.
This can be attributed to the sharp change in the slope of the potential in
the former model~\cite{Starobinsky:1980te,Martin:2011sn}.
Moreover, for the models of interest, the dip in the power 
spectrum occurs at the linear order of the perturbations. 
Though there may arise corrections to the power spectrum due to higher order 
correlations, we have checked, for instance, that the corrections due to the 
bispectrum are negligible for the parameters considered.


\noindent
\underline{\it Imprints on pre-ionization HI signal:}\/~We now turn to discuss 
the imprints of the inflationary power and bi-spectra with sharp dips that we 
have obtained on the~$21\,\mathrm{cm}$ signal of HI from the Dark Ages.
We briefly outline the essential points.
In the rest frame of a hydrogen atom, the hyperfine splitting of the ground 
state causes an energy difference that corresponds to the wavelength of 
$\lambda = 21.1\, \mathrm{cm}$. 
The spin temperature~$\Ts$ of this line is determined by three processes taking 
place in the early universe:~emission and absorption of CMB photons with a black 
body temperature~$\Tcmb$, collisions with atoms, and the mixing of the two levels 
caused by Ly-$\alpha$ photons (i.e. the Wouthuysen-Field effect).  
The spin temperature~$\Ts$ can be expressed in terms of~$\Tcmb$, the gas 
kinetic temperature~$\Tk$, and the color temperature of the Lyman-$\alpha$ 
photons~$T_{\alpha}$, as follows~\cite{Field1958,21cm_21cen}:~$\Ts
=(\Tcmb+\yc\, \Tk+y_{\alpha}\,T_{\alpha})/(1+\yc+y_{\alpha})$.
In this expression, $\yc$ and $y_\alpha$ determine the efficacy of the 
collisions between the hydrogen atoms and of the hydrogen atoms with the 
Lyman-$\alpha$ photons, respectively.
Note that, $\yc \propto n_{_{\mathrm{HI}}}$ and $y_\alpha \propto n_\alpha$,
where $n_{_{\mathrm{HI}}}$ and $n_\alpha$ denote the number density of HI
and the Lyman-alpha photons.
HI emits or absorbs $21\,\mathrm{cm}$ radiation from the CMB depending 
on whether $\Ts$ is greater than or less than~$\Tcmb$. 
This global temperature difference is observable and can be expressed 
as~\cite{21cm_21cen,2004ApJ...608..611G,Sethi05}
\begin{equation}
\Delta T_b(z) \simeq 30\, \l(1-\f{\Tcmb}{\Ts}\r)\,\l(\f{1+z}{10}\r)^{1/2}\,
\l(\f{\Omega_\mathrm{b}\, h^2}{0.022}\r)\, \mathrm{mK}.
\label{eq:overallnorm}
\end{equation}
The signal is observable at the frequency of $1420\, \mathrm{MHz}/(1+z)$
at a given redshift~$z$. 

Before the onset of the era of cosmic dawn, $y_\alpha=0$ and the dynamics
of $\Ts$ is entirely determined by the other two processes. 
For $z \gtrsim 50$, collisions dominate and hence $\Ts \simeq \Tk$. 
As $\Tk \simeq \Tcmb$ for $z\gtrsim 200$, $\Ts$ relaxes to $\Tcmb$ in this
redshift range and the observable signal is negligible. 
At lower redshifts ($z \lesssim 150$), $\Tk$ falls adiabatically 
as $1/a^2$ and, as $\Ts \simeq \Tk$, HI is observable in absorption. 
At even smaller redshifts, owing to the dilution of the gas, the
collisional coupling becomes progressively weaker and $\Ts$ relaxes 
to $\Tcmb$, causing the HI signal to diminish.
\enlargethispage*{0.75cm}

\begin{figure}[!t]
\includegraphics[width=1.0\linewidth]{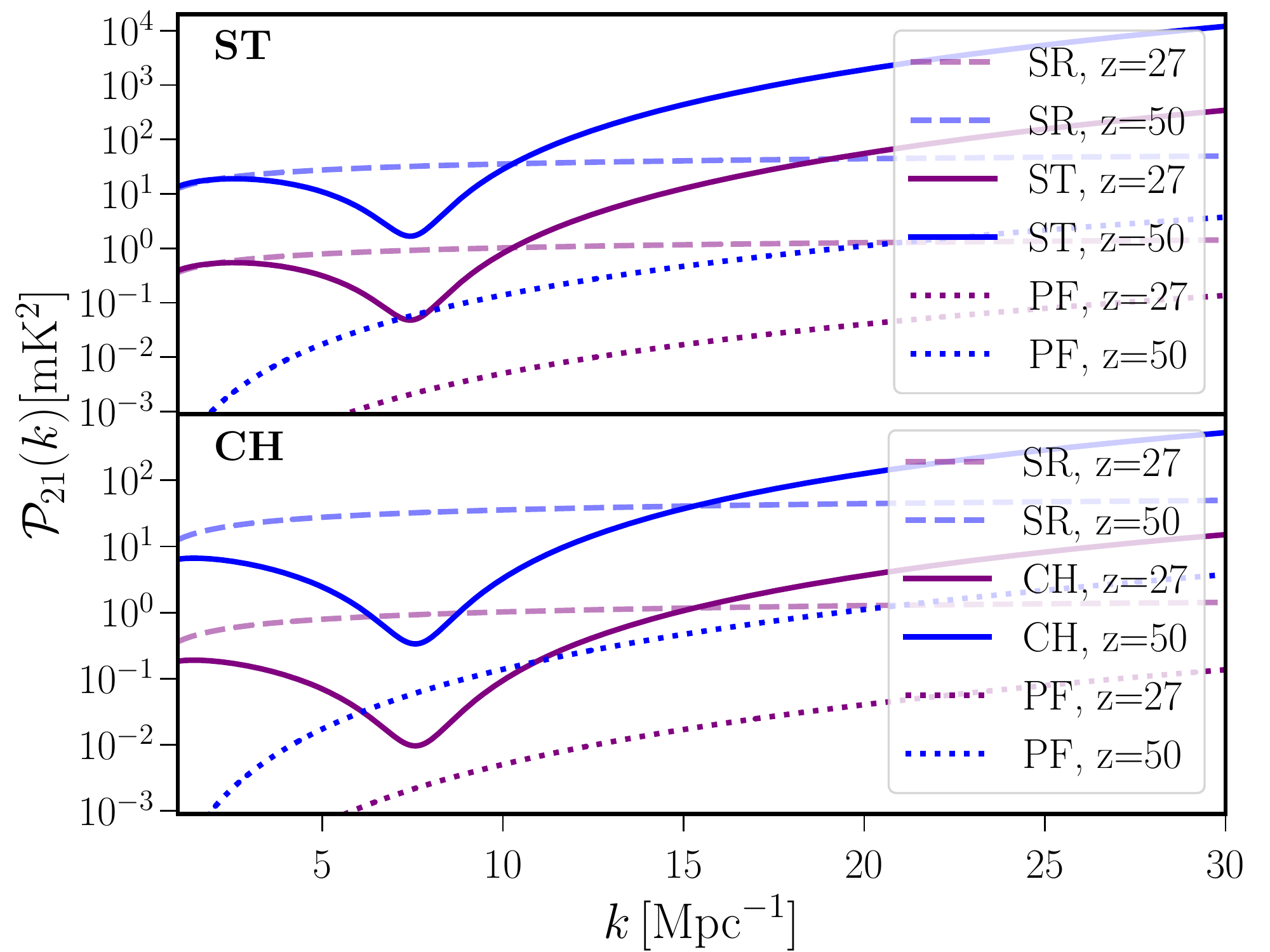}
\caption{The HI intensity power spectra arising from the ST and CH models
have been plotted at the redshifts of $z=27$ and $z=50$.
For comparison, we have also presented the HI intensity power spectra arising 
in a SR scenario leading to a nearly scale invariant, power law primordial 
scalar power spectrum.
We have also included the power spectra due to PF at the corresponding 
redshifts.}\label{fig:p21}
\end{figure}

\begin{figure}[!t]
\includegraphics[width=1.0\linewidth]{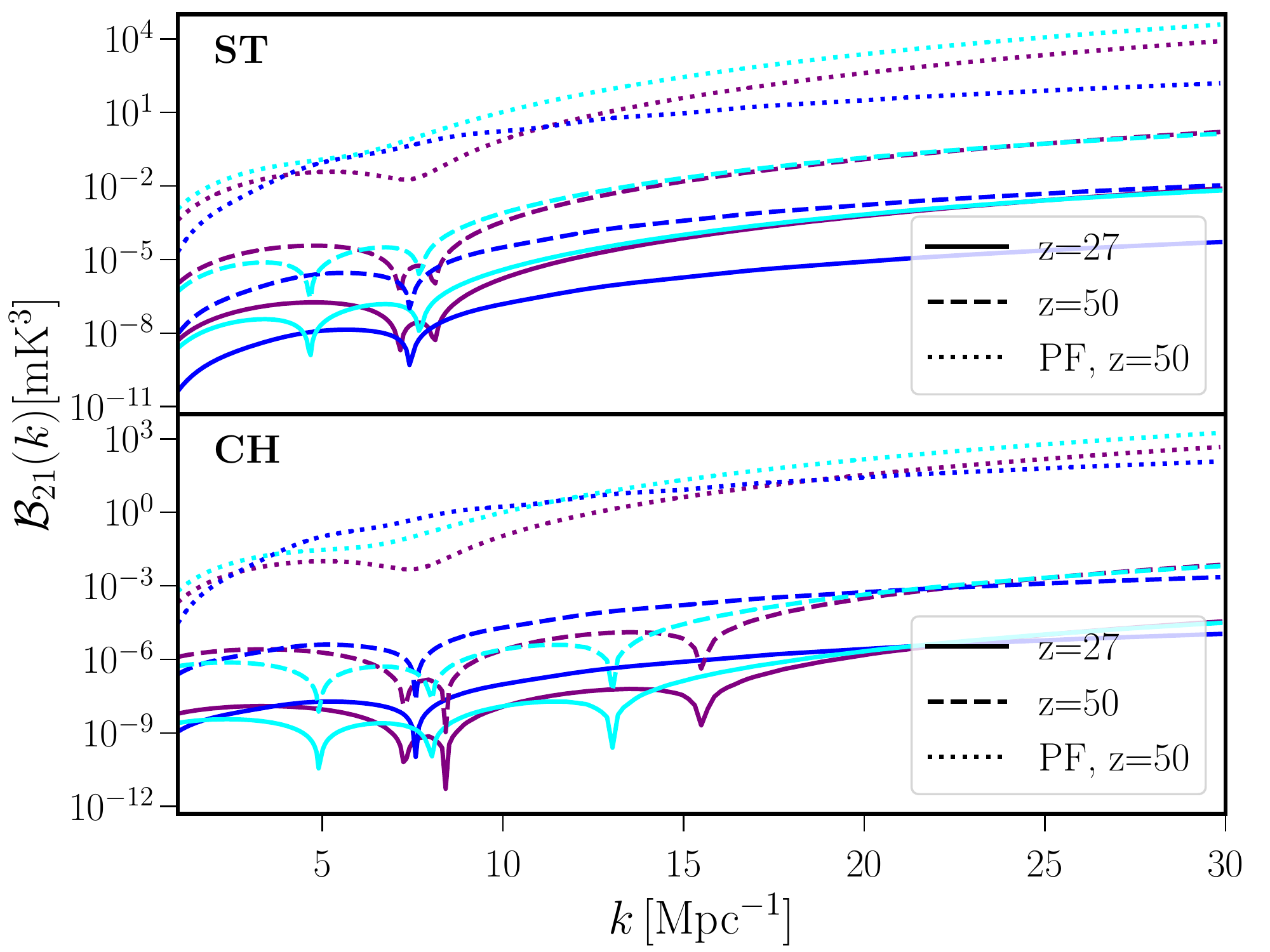}
\caption{The HI intensity bispectra  arising from the ST and CH models have been 
presented in the equilateral, squeezed and the flattened limits with the same
choice of colors as in Fig.~\ref{fig:sbs}.
We have focused around $\kdip$ and we have set $k_\mathrm{sq}=k/100$ to arrive 
at the behavior in the squeezed limit. 
The associated PF have also been indicated.}\label{fig:b21}
\end{figure}

We have relegated the details of the computation of the power 
and bi-spectra of the HI signal to App.~II.
In Fig.~\ref{fig:p21}, we present the HI intensity power spectrum 
arising in the ST and CH models at the redshifts of $z=27$ and $z=50$.
In the figure, we have also included the results from a typical SR model, 
along with the contribution from PF.
In Fig.~\ref{fig:b21}, we have illustrated the HI intensity bispectrum arising 
in the ST and CH models at two redshifts in the equilateral, squeezed and 
flattened limits, along with the contribution from PF. 
\enlargethispage*{1.0cm}

Our main findings, shown in Figs.~\ref{fig:p21} and~\ref{fig:b21}, clearly indicate
that, in the presence of an epoch of USR, there arises a significant dip in the HI 
intensity power and bi-spectra over the scales $1 \lesssim k \lesssim 10\, \mpcinv$,
when compared to a typical SR scenario. 
The HI signal arising from the inflationary bispectrum is seen to be smaller than 
the contribution from PF by many orders of magnitude. 
As the Poisson contribution to the bispectrum depends on the HI power spectrum [cf. 
Eq.~\eqref{eq:hibpo}], the detection of this signal would provide further evidence 
of the presence of a null or a dip in the inflationary power spectrum. 


\noindent
\underline{\it Sensitivity:}\/~We now explore the feasibility of the detection
of the HI intensity power spectrum over scales of interest. 
As can be seen in Fig.~\ref{fig:p21}, our main predictions are over the scales $1 
\lesssim k \lesssim 100 \, \mpcinv$.
The signal strength at such scales is  of the order of $10$--$1000\, (\mathrm{mK})^2$ 
in the frequency range $25$--$50\, \mathrm{MHz}$, for the redshift range $z \simeq 
25$--$50$. 
While the signal at $z\simeq 25$ is accessible to SKA-Low (see, 
e.g., Ref.~\cite{Koopmans:2015sua}), we expect the signal at $z\simeq 50$ to be
more pristine (i.e. less contaminated by astrophysical processes close to the era 
of cosmic dawn) and dominant.
Such a signal could be explored by planned lunar 
missions~\cite{Furlanetto:2019jso,Cole:2019zhu,Koopmans:2019wbn}. 
Under suitable assumptions (for a detailed discussion and methodology, see 
the supplementary material, which includes 
Refs.~\cite{Paul:2016blh,wayth2018phase,Furlanetto:2006wp}), the brightness 
temperature sensitivity of $1$--$10 \, 
(\mathrm{mK})^2$ can be achieved for the scales of interest.
A comparison with Fig.~\ref{fig:p21} immediately suggests that the attainable 
sensitivity should allow the detection of the dip due to the epoch of USR in 
the HI power spectrum. 


\noindent
\underline{\it Conclusions:}\/~In models of inflation driven by a single, canonical 
scalar field, an extended phase of USR leads to a {\it null}\/ in {\it all}\/ the 
primordial correlations at a specific wavenumber~$\kdip$.
We have considered scenarios in which the null in the primordial correlations 
occurs over wavenumbers $1 \lesssim \kdip \lesssim 10\,\mathrm{Mpc}^{-1}$. 
We show that future experiments should have the sensitivity to detect a damping 
of power in the HI signal due to the nulls at the level of the power and bi-spectra.


\noindent
\underline{\it Acknowledgements:}\/~SB thanks Yi-Peng Wu for helpful discussions.
HVR acknowledges support from the Indian Institute of Science Education and 
Research Kolkata through postdoctoral fellowship.
LS acknowledges support from the Science and Engineering Research Board, Department 
of Science and Technology, Government of India, through the Core Research 
Grant~CRG/2018/002200.
SB is supported by funding from the European Union’s Horizon 2020 research and 
innovation programme under grant agreement No.~101002846 (ERC CoG ``CosmoChart'') 
as well as support from the Initiative Physique des Infinis (IPI), a research 
training program of the Idex SUPER at Sorbonne Universit\'{e}.


\appendix
\section{Appendix~I:~Inflationary scalar bispectrum}\label{app:SBS}

\begin{figure}
\centering
\includegraphics[width=1.0\linewidth]{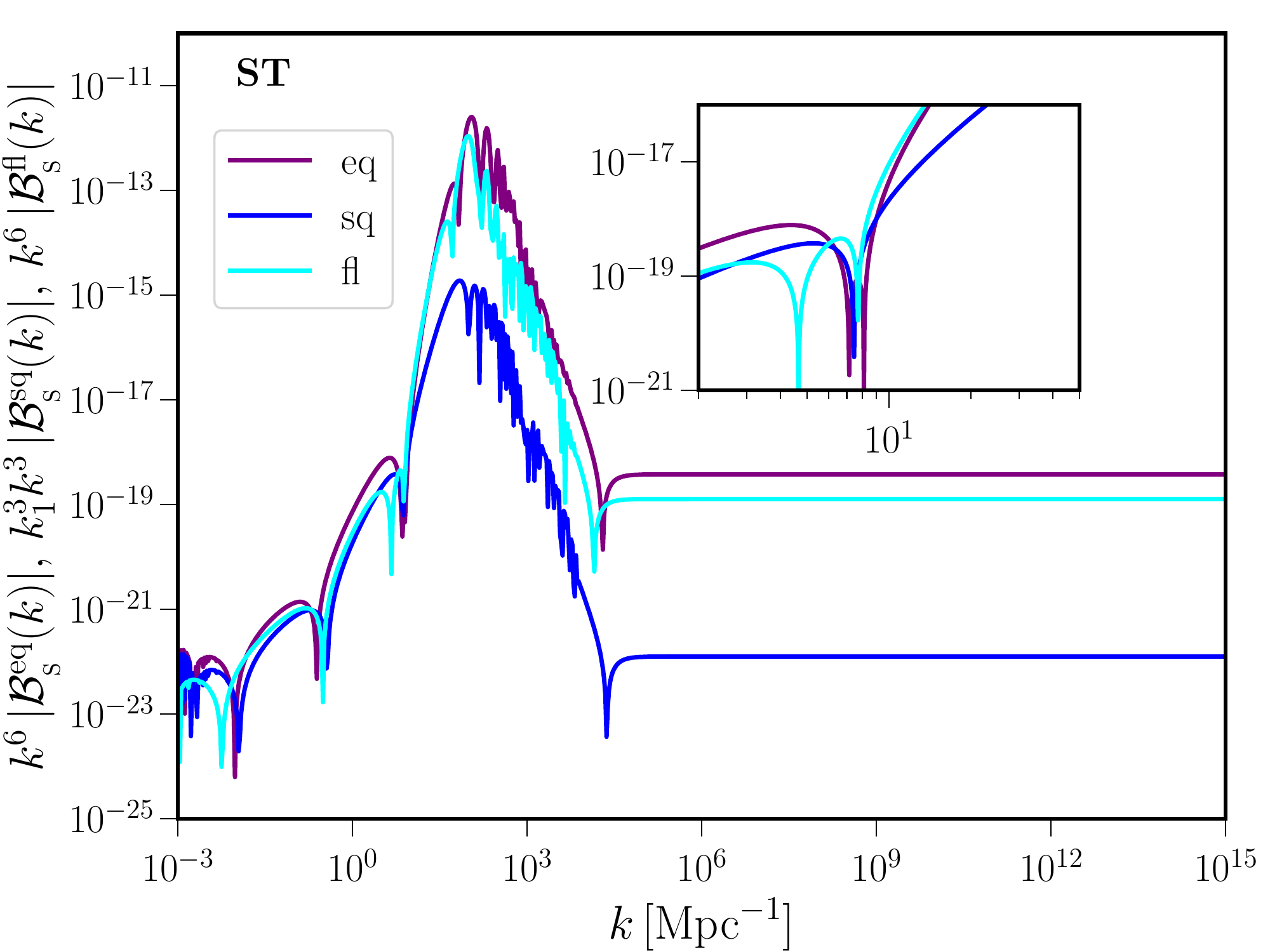}
\includegraphics[width=1.0\linewidth]{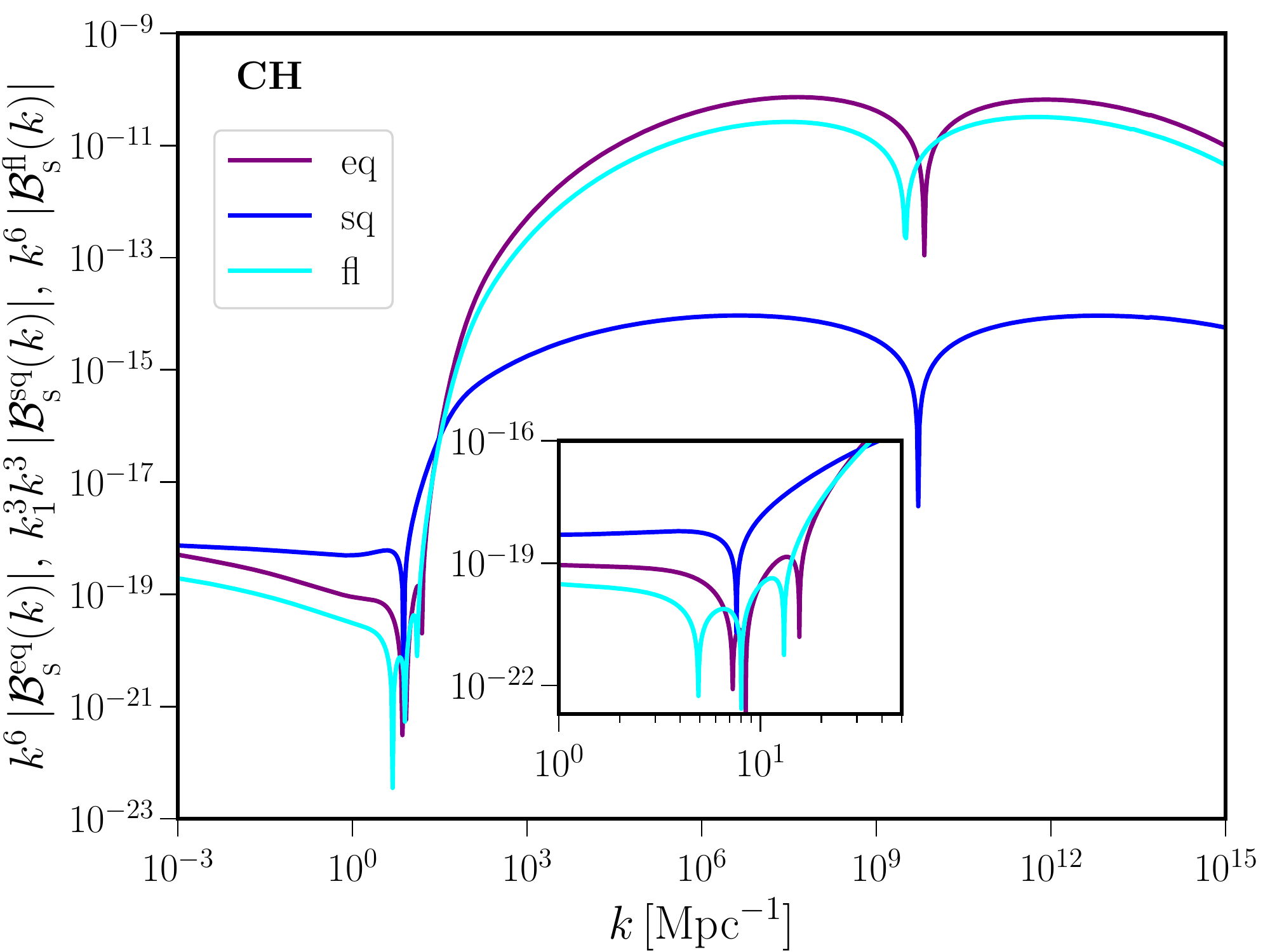}
\caption{The scalar bispectra arising in the ST and CH models have been 
illustrated in the equilateral (eq), squeezed (sq) and flattened (fl) limits.
As highlighted in the insets, the bispectra also exhibit a sharp dip at the 
same location (i.e. at $7.6\,\mpcinv$) as the power spectra in the previous 
figure.}\label{fig:sbs}
\end{figure}

In this appendix, we have plotted the bispectra in the equilateral (i.e. when $k_1=k_2=k_3$),  
squeezed (when $\vka \to 0$ and $\vkb = -\vkc = \vk$) and flattened (when
$k_1=k_2=k$ and $k_3=2\,k$) limits.
In Fig.~\ref{fig:sbs}, we have illustrated the dimensionless quantities
such as $k^6\,\bs(k)$ in the equilateral and flattened limits, and $k_1^3\,
k^3\,\bs(k)$ in the squeezed limit.
We find that a dip in the bispectra arises in {\it all}\/ the limits at the
same location of the dip (viz. at $\kdip=7.6\,\mpcinv$) in the power spectra.


\section{Appendix~II:~Computation of power and bi-spectra of $21$ cm signal}
\label{app:HI-spectra}

In this appendix, we provide the details of the computation of power and 
bi-spectra of the HI intensity signal in terms of the primordial spectra.

At linear order in the perturbations, the density fluctuations in HI
follow the baryonic perturbations.
This allows us to express the fluctuating component of the HI signal in 
the pre-reionization epoch as $\delta T({\bm x}) = \Delta T_b(z)\, 
\delta_{_{\mathrm{HI}}}({\bm x})$, where $\delta_{_{\mathrm{HI}}}({\bm x})$ 
denotes the inhomogeneities in the density of the neutral gas. 
We shall ignore the redshift-space distortions in our discussion.
At small scales, these perturbations are wiped out due to acoustic damping 
in the pre-recombination era and are re-generated by dark matter potential 
wells in the post-recombination era. 
In linear theory, the baryonic perturbations can be expressed in terms of the 
inflationary scalar power spectrum as $\mathcal{P}_{_{\mathrm{HI}}}(k,z)
= T^2(k,z)\, \ps(k)$, where $T(k,z)$ is the transfer function for baryons 
defined such that $T(k,z)$ tends to unity for small $k$ (see, for instance, Ref.~\cite{2003moco.book.....D}). 
This allows us to write the HI intensity power spectrum at a redshift~$z$ in
$(\mathrm{mK})^2$ as follows:
\begin{equation}
\mathcal{P}_{21}(k,z) = \l[\Delta T_b(z)\r]^2\, T^2(k,z)\, \ps(k).
\end{equation}

At any redshift, a fraction of baryons, say, $f_c$, collapse to form halos. 
The baryons in these halos remain neutral, since for the parameters of interest, 
the masses of collapsed halos are $\mathcal{O}(10^6\, \mathrm{M}_\odot)$ and 
the virial temperature of these halos is less than $1000\,\mathrm{K}$, too 
small to ionize the gas via collisional processes (we evade the free-free 
constraints on the excess matter power discussed in Ref.~\cite{2022PhRvD.105f3531A}; 
also see Ref.~\cite{2020MNRAS.497..279C}). 
The HI intensity power spectrum from PF at a given redshift in $\rm (mK)^2$ 
is given by (in this context, see, for instance, 
Ref.~\cite{1980lssu.book.....P})
\begin{equation}
\mathcal{P}_{21}^{^{\mathrm{PF}}}(k,z) 
= \l[f_c\, \Delta T_b(z)\r]^2\, \f{k^3}{2\,\pi^2}\,\f{1}{\bar{n}},
\end{equation}
where $\bar{n}$ is the mean comoving number density of halos. 
The collapsed fraction $f_c$ and the number density of halos $\bar{n}$ at any
redshift  can be computed using the Press-Schecheter formalism. 
For instance, in the CH model, at $z=50$, $f_c = 0.17$ and $\bar{n} = 18754\, 
\mathrm{Mpc}^{-3}$, while at $z=100$, $f_c = 7\times 10^{-3}$ and $\bar{n} = 1549\, 
\mathrm{Mpc}^{-3}$.
At both redshifts, the mass function is dominated by halos of mass  
$M \lesssim 5 \times 10^5\, \mathrm{M}_\odot$. 
At $z=27$, $\bar{n} = 27638\, \mathrm{Mpc}^{-3}$ and $f_c = 0.45$, with 
halos of $M \lesssim 2 \times 10^6 \, \mathrm{M}_\odot$ making the most 
significant contribution to the mass function.
We have plotted the spectra $\mathcal{P}_{21}(k,z)$ and 
$\mathcal{P}_{21}^{^{\mathrm{PF}}}(k,z)$ for the ST and CH models in 
Fig.~\ref{fig:p21}.

Given the scalar bispectrum~$\bs(\vka,\vkb,\vkc)$ generated during inflation, 
the HI intensity bispectrum at any redshift can be expressed in units of 
$(\mathrm{mK})^3$ as 
\begin{eqnarray}
{\cal B}_{21}({\bm k}_1,{\bm k}_2,{\bm k}_3,z)  
&=& \f{\l[\Delta T_b(z)\r]^3}{2\,\pi^2}\,
T(k_1,z)\, T(k_2,z)\, T(k_3,z)\nn\\
& &\times\, \f{k_1^3\,k_2^3\,k_3^3}{(k_1^3+k_2^3+k_3^3)}\,
\bs(\vka,\vkb,\vkc).\qquad
\end{eqnarray}
Also, the bispectrum from PF of discrete sources is given by (see e.g., Ref.~\cite{1980lssu.book.....P})
\begin{eqnarray}
{\cal B}_{21}^{\rm PF}({\bm k}_1,{\bm k}_2,{\bm k}_3,z)  
& = & \f{f_c^3\, k_1^3\,k_2^3\,k_3^3\,\Delta T_b(z)}{(k_1^3+k_2^3+k_3^3)\,\bar{n}}\,
\Biggl\{-\f{2\, [\Delta T_b(z)]^2}{\bar{n}}\nn\\
& &+\,\f{\mathcal{P}_{21}(k_1)}{k_1^3} + \f{\mathcal{P}_{21}(k_2)}{k_2^3} 
+ \f{\mathcal{P}_{21}(k_3)}{k_3^3}\Biggr\}.\nn\\
\label{eq:hibpo}
\end{eqnarray}
We have presented the HI intensity bispectrum at redshifts of 
$z=27$ and $50$ in Fig.~\ref{fig:b21} along with the corresponding
PF computed at $z=50$.


\input{supplement}

\end{document}

%% file: supplement.tex
\clearpage

\section{Supplemental material for\\
`Observing nulling of primordial correlations via the $21\, \mathrm{cm}$ signal'}

In the following sections, we shall gather material that supplement 
the discussions in the manuscript.


\subsection{Starobinsky functions}\label{sec:starobinksyfunctions}

The scalar power spectrum in the ST model can be analytically calculated 
to be~\cite{Starobinsky:1992ts,Martin:2011sn,Sreenath:2014nka}
\begin{eqnarray}
\ps(k) &=& \ps^0\,\l(\f{A_+}{A_-}\r)^2\,\biggl[I(k) 
+ I_c(k)\,\cos{\l(\f{2\,k}{k_0}\r)}\nn\\ 
& &+\, I_s(k)\,\sin{\l(\f{2\,k}{k_0}\r)}\biggr],\label{eq:ps-smii}
\end{eqnarray}
where $k_0$ is the wavenumber that leaves the Hubble radius when $\phi=\phi_0$,
while the quantities $\ps^0$ and $\epsilon_{1_-}$ are related to the parameters 
of the potential by the relations
\begin{equation}
\ps^0 \simeq \f{V_0}{24\,\pi^2\,\Mpl^4\,\epsilon_{1_+}},\quad
\epsilon_{1_+} \simeq \f{\Mpl^2}{2}\l(\f{A_+}{V_0}\r)^2.
\end{equation}
The functions $I(k)$, $I_c(k)$ and $I_s(k)$ that appear in the power spectrum
above are given by
\begin{subequations}
\begin{eqnarray}
I(k) &=& 1 + \f{9}{2}\,\l(\f{\Delta A}{A_+}\r)^2\,\l(\f{k_0}{k}\r)^2\nn\\
& &\times\,\l[1 + 2\,\l(\f{k_0}{k}\r)^2 + \l(\f{k_0}{k}\r)^4\r],\\
I_c(k) &=& \f{3}{2}\,\f{\Delta A}{A_+}\,\l(\f{k_0}{k}\r)^2 
\l[3\,\f{\Delta A}{A_+} - 4 - 3\,\f{\Delta A}{A_+}\,\l(\f{k_0}{k}\r)^4\r],\nn\\
\\
I_s(k) &=& -3\,\f{\Delta A}{A_+}\,\f{k_0}{k}\nn\\
& &\times\,\biggl[1 + \l(\f{3\,\Delta A}{A_+} - 1\r)\,\l(\f{k_0}{k}\r)^2
+ 3\,\f{\Delta A}{A_+}\,\l(\f{k_0}{k}\r)^4\biggr],\nn\\
\end{eqnarray}
\end{subequations}
where $\Delta A = A_- - A_+$. 
Actually, we need to smoothen the potential to be able to compute the spectrum 
numerically (in this context, see Refs.~\cite{Martin:2011sn,Sreenath:2014nka,
Ragavendra:2020old}; also see the discussion that follows).
We find that the above analytical form for the power spectrum matches the numerical
result very well.


\subsection{Parameters and background evolution}

In the case of the ST model, for numerical analysis, we smoothen the 
potential [cf.~Eq.~(4) of the manuscript]
and work with the potential given 
by~\cite{Martin:2014kja,Ragavendra:2020old}
\begin{eqnarray}
V(\phi) 
&=& V_0 + \f{1}{2}\,\l(A_{+} + A_{-}\r)\,\l(\phi-\phi_0\r)\nn\\ 
& &+\, \f{1}{2}\,\l(A_{+} - A_{-}\r)\,\l(\phi-\phi_0\r)\,
\tanh{\l(\f{\phi-\phi_0}{\Delta\phi}\r)}.\qquad\label{eq:sm2}
\end{eqnarray}
We set $V_0 = 2.48 \times10^{-11}\,\Mpl^4$ and $A_+ = V_0/(100\,\Mpl)$ to 
achieve COBE  normalization over the CMB scales. 
We choose $A_-=V_0/(4\times10^3\Mpl)$ and $\phi_0=0.5628\,\Mpl$.
If we choose the initial value of the field and the first SR parameter to be 
$\phi_\mathrm{i}=0.84348\,\Mpl$ and $\epsilon_{1\mathrm{i}}=4.0\times 10^{-5}$,
respectively, we find that the pivot scale leaves the Hubble radius around 
$6.4$ e-folds before field crosses~$\phi_0$.
These choices for the parameters leads to a rise in power on smaller scales 
that is consistent with the FIRAS constraints on
$\mu$-distortions~\cite{Gow:2020bzo}. 

In the CH model, we choose the values of the parameters to be $V_0 = 7.05\times 
10^{-8}\,\Mpl^4$ and $\phi_0=1\,\Mpl$.
The dimensionless parameters $a$ and $b$ are related to $c$ and the location of 
the point of inflection, say, $z_\mathrm{c}$.
Choosing the parameters $(c, z_\mathrm{c}) = (2.850, 0.820)$ leads to $(a, b)=
(1.694, 0.601)$.
The value of the parameter~$b$ is then fine tuned to achieve the desired duration 
of the USR epoch so that the rise in power is consistent with the constraints 
from FIRAS~\cite{Gow:2020bzo}.
For these values of the model parameters and with the initial value of field 
$\phi_\mathrm{i} = 5.55\,\Mpl$ and the first SR parameter $\epsilon_{1\mathrm{i}}
=10^{-3}$, we achieve about $70$~e-folds of inflation.
We set the pivot scale to exit the Hubble radius at $46.5$ e-folds before the 
end of inflation. 
The epoch of USR begins as the field crosses the point $\phi=0.82\,\Mpl$, around 
$40$~e-folds before the end of inflation.


\subsection{Reconstructing a model with a phase of USR}

\begin{figure*}
\centering
\includegraphics[width=0.475\linewidth]{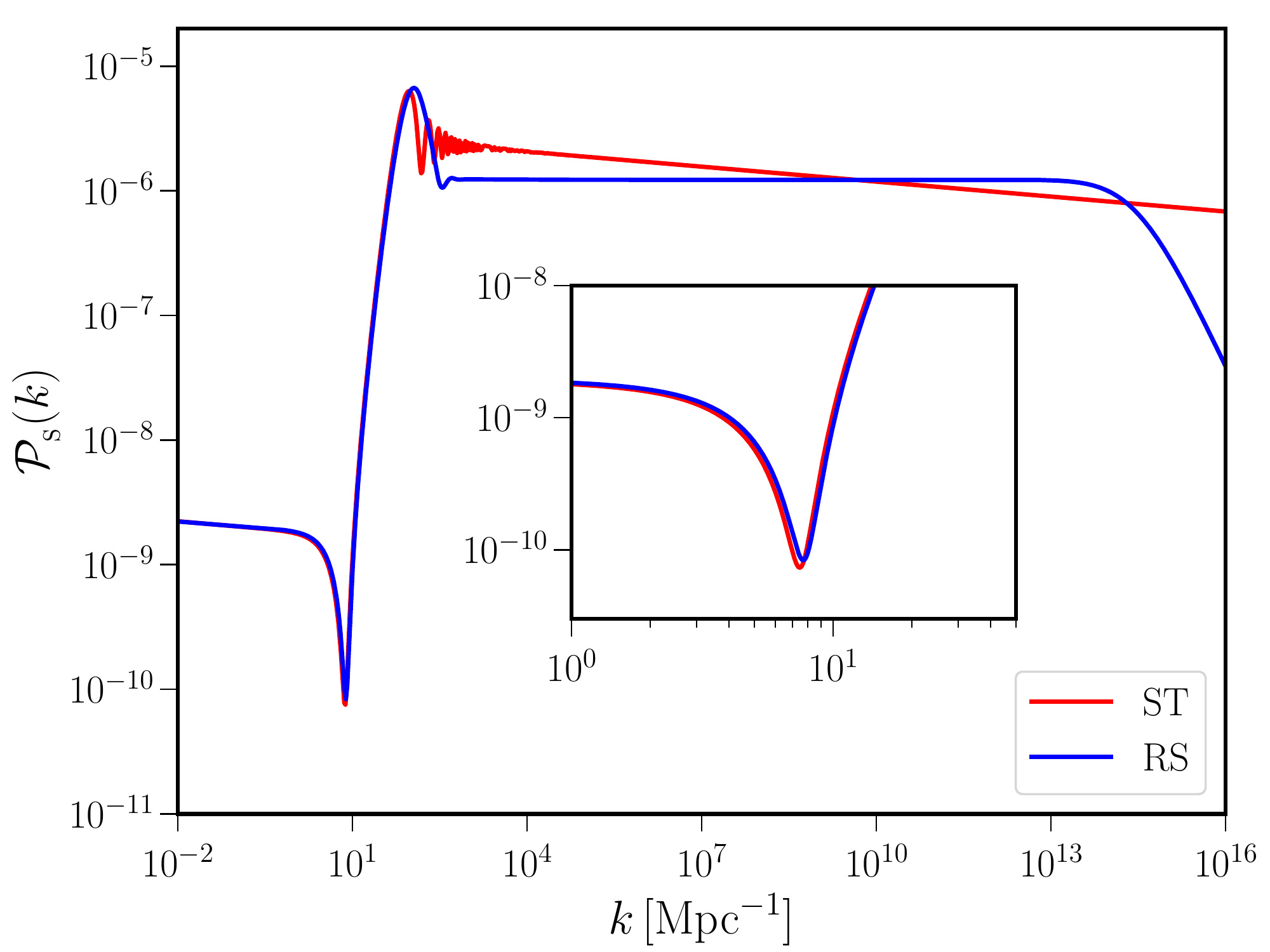}
\hskip 5pt
\includegraphics[width=0.475\linewidth]{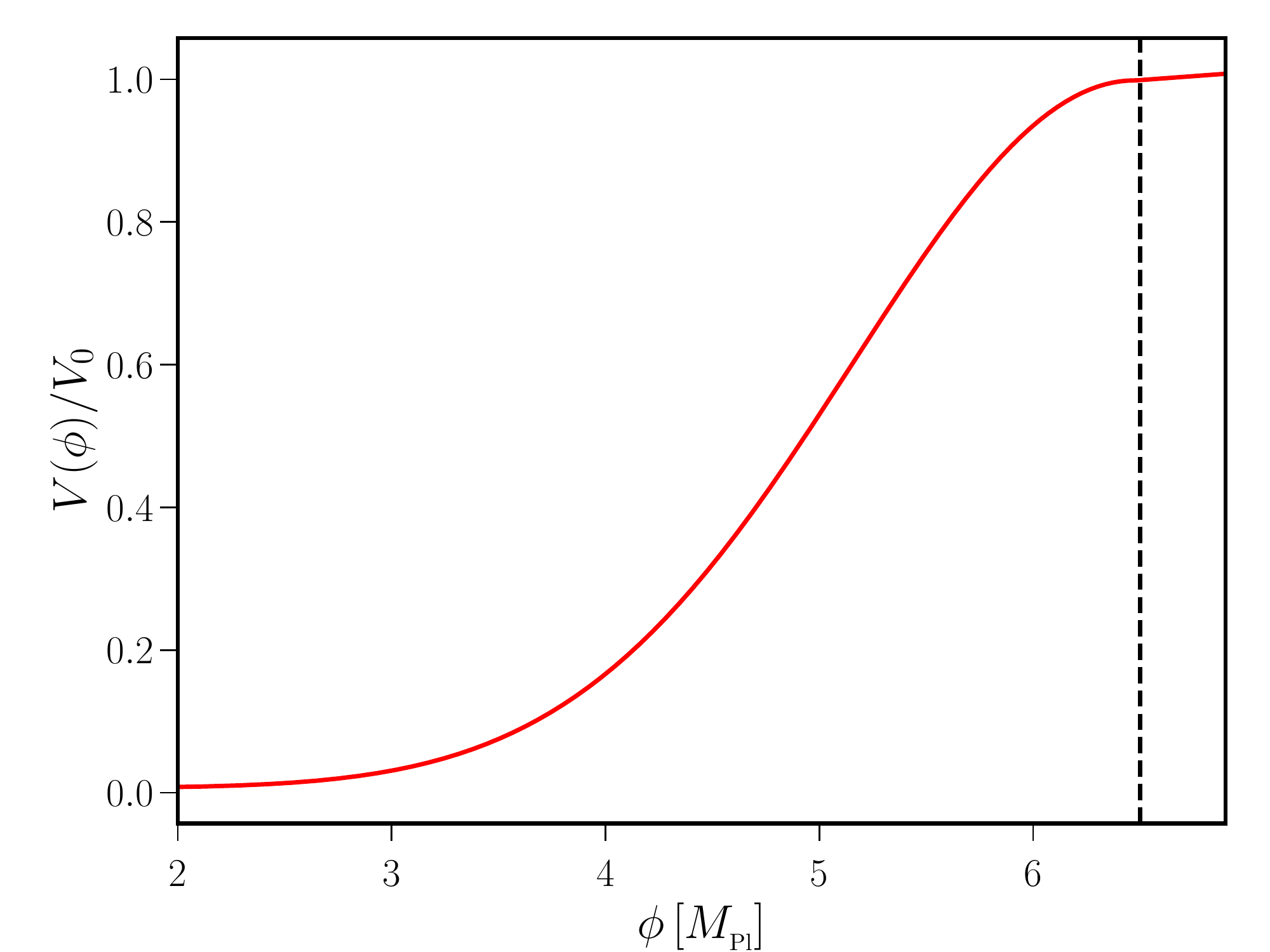}
\caption{The scalar power spectrum $\ps(k)$ arising from the RS involving 
an epoch of USR has been presented along with the power spectrum from the 
ST model (on the left).
Notice that both the power spectra have identical shapes close to~$\kdip$
(as also highlighted in the inset).
In the RS, the power spectrum decays at large wavenumbers due to the 
termination of inflation.
We have also illustrated the form of the potential $V(\phi)$ describing the 
RS (on the right).
Interestingly, to the extent permitted by numerical analysis, we find that 
the reconstructed potential too contains a point of inflection around $\phi
=6.5\,\Mpl$ (indicated by the black vertical line). 
The parameter $V_0$ indicated has been taken to be $3\,H_\mathrm{i}^2\,\Mpl^2$,
where $H_\mathrm{i}$ is the initial value of the Hubble parameter.
We have chosen $H_\mathrm{i}=6.98\times10^{-6}\,\Mpl$ so as to achieve COBE 
normalization of the scalar power spectrum over the CMB scales.}\label{fig:ps-eps1}
\end{figure*}

We had pointed out that, at the pivot scale of $0.05\,\mpcinv$, the ST and 
CH models lead to a scalar spectral index that is quite removed from the 
mean value of $\ns=0.96$ suggested by Planck~\cite{Aghanim:2018eyx}.
We had also mentioned that, it is challenging to arrive at a model involving
a single, canonical scalar field that leads to a significant enhancement of 
power on small scales and is consistent with the constraints from Planck on 
the CMB scales~\cite{Iacconi:2021ltm,Karam:2022nym}.
To overcome this issue, one may resort to the reconstruction of a scenario 
of inflation using a parametric modeling of the first SR parameter $\epsilon_1$.
We find that the following functional form of $\epsilon_1(N)$ leads to the desired
behavior of the background and power spectra (for a detailed discussion, see
Ref.~\cite{Ragavendra:2020sop}):
\begin{eqnarray}
\epsilon_1(N) &=& \epsilon_{1 a}\,(1+\epsilon_{2 a}\,N)
\l[1 - \tanh \l(\f{N-N_1}{\Delta N_1} \r)\r] + \epsilon_{1b}\nn\\
& & + \exp\left(\f{N-N_2}{\Delta N_2}\right),
\end{eqnarray}
where the quantities $(\epsilon_{1 a},\epsilon_{2 a},\epsilon_{1 b},N_1,N_2,
\Delta N_1, \Delta N_2)$ are parameters which we shall choose suitably.
The parameters $(\epsilon_{1a},\epsilon_{2a})$ help to achieve appropriate 
values for the scalar spectral index and the tensor-to-scalar ratio over the CMB 
scales.
While the parameters $(N_1, \Delta N_1, \epsilon_{1 a},\epsilon_{1 b})$ control 
the onset and duration of the epoch of USR, the parameters $(N_2,\Delta N_2)$ 
regulate the behavior of the background after the USR phase until the termination
of inflation.
We have chosen the values of these parameters such that the dip in the power 
spectrum arises at the desired location and inflation is terminated after 
$60$ e-folds. 
The values of the parameters we have worked with are as follows: 
$(\epsilon_{1\rm a},\epsilon_{2 \rm a}, N_1, \Delta N_1, 
\epsilon_{1 b},N_2\Delta N_2)= (10^{-4},0.05,16.5,0.3,5\times10^{-7},60,1)$.
In Fig.~\ref{fig:ps-eps1}, we have illustrated the potential arrived at numerically 
from the above form of the first SR parameter (i.e. the reconstructed scenario, RS) 
as well as the resulting scalar power spectrum.
Note that the resulting scalar power spectrum from the RS closely resembles
the spectrum in the ST model around~$\kdip$.


\subsection{Computation of scalar power and bi-spectrum}

Let $\cR_{\bm k}$ denote the Fourier modes associated with the curvature 
perturbation~$\cR$.
Recall that the scalar power spectrum $\ps(k)$ is defined through the 
relation
\begin{equation}
\langle \hat{\cR}_{\vk}(\ee)\, 
\hat{\cR}_{\vk'}(\ee)\rangle=\f{(2\,\pi)^2}{2\,k^3}\,\ps(k)\,     
\delta^{(3)}(\vk+\vk'),
\end{equation}
where $\ee$ denotes the conformal time close to the end of inflation and the
expectation value on the left hand side is to be taken in the Bunch-Davies 
vacuum.
In terms of the positive frequency Fourier modes that describe the curvature 
perturbations, say, $f_k$, the power spectrum is given by
\begin{equation}
\ps(k)=\f{k^3}{2\,\pi^2}\,\vert f_k\vert^2.    
\end{equation}

The primordial scalar bispectrum in Fourier space, denoted 
as $\bs(\vka,\vkb,\vkc)$, is defined through the relation
\begin{eqnarray}
\langle \hat{\cR}_{\vka}(\ee)\, 
\hat{\cR}_{\vkb}(\ee)\, \hat{\cR}_{\vkc}(\ee)\rangle 
&=& (2\,\pi)^3\, \bs(\vka,\vkb,\vkc) \nn \\
& &\times\, \delta^{(3)}(\vka+\vkb+\vkc),\qquad\;\;\label{eq:g-def}
\end{eqnarray}
with the expectation value on the left hand side to be evaluated in the 
perturbative vacuum~\cite{Maldacena:2002vr,Seery:2005wm,Chen:2010xka}.
The bispectrum $\bs(\vka, \vkb, \vkc)$ can be calculated using the third order 
action governing the perturbations and the standard rules of perturbative quantum 
field theory~\cite{Martin:2011sn,Arroja:2011yj}.
It can be shown that, in the situations of our interest, the scalar bispectrum 
can be expressed as (see, for instance,
Refs.~\cite{Martin:2011sn,Hazra:2012yn,Ragavendra:2020sop,Ragavendra:2020sop})
\begin{eqnarray}
\bs(\vka,\vkb,\vkc) 
&=& \f{\Mpl^2}{(2\,\pi)^{9/2}}\, 
\sum_{C=1}^{6} \biggl[f_{k_1}(\ee)\, f_{k_2}(\ee)\,f_{k_3}(\ee)\nn\\
& & \times\, B_{_{C}}(\vka,\vkb,\vkc) +\,\mathrm{complex~conjugate}\biggr]\nn\\
& & +\, \f{1}{(2\,\pi)^{9/2}}\,\mathcal{B}_{7}(\vka,\vkb,\vkc),\label{eq:sbs}
\end{eqnarray}
where $f_k$ are the positive frequency Fourier modes associated with the curvature 
perturbation we mentioned above.
The quantities $B_{_{C}}(\vka,\vkb,\vkc)$ represent six integrals that 
involve the scale factor, the slow roll parameters, the mode functions~$f_k$ 
and their time derivatives~$f_k'$.
They correspond to the six bulk terms appearing in the cubic order action 
and are described by the following expressions
\begin{subequations}\label{eq:cG}
\begin{eqnarray}
B_1(\vka,\vkb,\vkc)
&=& 2\,i\,\int_{\ei}^{\ee} \d\eta\; a^2\, 
\epsilon_{1}^2\, \biggl(f_{k_1}^{\ast}\,f_{k_2}'^{\ast}\,
f_{k_3}'^{\ast}\nn\\
& &+\,\mathrm{two~permutations}\biggr),\label{eq:cG1}\\
B_2(\vka,\vkb,\vkc)
&=&-2\,i\;\l(\vka\cdot \vkb +\,{\rm two~permutations}\r)\nn\\ 
& &\times\,\int_{\ei}^{\ee} \d\eta\; a^2\, 
\epsilon_{1}^2\, f_{k_1}^{\ast}\,f_{k_2}^{\ast}\,
f_{k_3}^{\ast},\label{eq:cG2}\\
B_3(\vka,\vkb,\vkc)
&=&-2\,i\,\int_{\ei}^{\ee} \d\eta\; a^2\,\epsilon_{1}^2\,
\biggl(\f{\vka\cdot\vkb}{k_2^2}\,
f_{k_1}^{\ast}\,f_{k_2}'^{\ast}\, f_{k_3}'^{\ast}\nn\\
& &+\,\mathrm{five~permutations}\biggr),\label{eq:cG3}\\
B_4(\vka,\vkb,\vkc)
&=& i\,\int_{\ei}^{\ee} \d\eta\; a^2\,\epsilon_{1}\,\epsilon_{2}'\, 
\biggl(f_{k_1}^{\ast}\,f_{k_2}^{\ast}\,f_{k_3}'^{\ast}\nn\\
& &+\, \mathrm{two~permutations}\biggr),\label{eq:cG4}\\
B_5(\vka,\vkb,\vkc)
&=&\frac{i}{2}\,\int_{\ei}^{\ee} \d\eta\; 
a^2\, \epsilon_{1}^{3}\;\biggl(\f{\vka\cdot\vkb}{k_2^2}\,
f_{k_1}^{\ast}\,f_{k_2}'^{\ast}\, f_{k_3}'^{\ast}\nn\\
& &+\, \mathrm{five~permutations}\biggr),\label{eq:cG5}\\
B_6(\vka,\vkb,\vkc) 
&=&\frac{i}{2}\,\int_{\ei}^{\ee}\d\eta\, a^2\, \epsilon_{1}^{3}\,
\biggl[\f{k_1^2\,\l(\vkb\cdot\vkc\r)}{k_2^2\,k_3^2}\, 
f_{k_1}^{\ast}\, f_{k_2}'^{\ast}\, f_{k_3}'^{\ast}\nn\\ 
& &+\, \mathrm{two~permutations}\biggr].\label{eq:cG6}
\end{eqnarray}
\end{subequations}
The term $\mathcal{B}_{7}(\vka,\vkb,\vkc)$ corresponds to a temporal
boundary term in the cubic order action and is given by
\begin{eqnarray}
\mathcal{B}_{7}(\vka,\vkb,\vkc)
&=& -i\,\Mpl^2\,\l[f_{k_1}(\ee)\,f_{k_2}(\ee)\,f_{k_3}(\ee)\r]\nn\\
& &\times\, \biggl[a^2\epsilon_1\epsilon_{2}\,
f_{k_1}^{\ast}(\eta)\,f_{k_2}^{\ast}(\eta)\,f_{k_3}'^{\ast}(\eta)\nn\\ 
& &+\, \mathrm{two~permutations} \biggr]_{\eta_i}^{\ee}\nn\\
& &+\,\mathrm{complex~conjugate}.\label{eq:G7}
\end{eqnarray} 

In SR scenarios, typically, one evolves the perturbations from well inside 
the Hubble radius (say, when $k\simeq 10^2\,a\,H$) to adequately late times 
when the modes are on super-Hubble scales (say, when $k\simeq 10^{-5}\, a\,
H)$, to arrive at the power spectrum~\cite{Hazra:2012yn}.
When there occur departures from SR, in particular, in the presence of an 
epoch of USR, it becomes important to evolve the perturbations until very
late times, close to the end of inflation~\cite{Ragavendra:2020old,    
Ragavendra:2020sop}.
These arguments broadly apply for the computation of the scalar bispectrum 
as well.
However, note that, apart from numerically evaluating the mode functions,
to obtain the bispectrum, we also need to carry out integrals over the mode
functions, their time derivatives and the background quantities.
Since the bispectrum, in general, involves three wavenumbers, we need to
evolve the perturbations and carry out the integrals from the time when 
the smallest of the three wavenumbers of interest is well inside the Hubble 
radius to a time close to the end of inflation (for further details, see 
Ref.~\cite{Ragavendra:2020sop}).
It is these procedures that we have adopted to arrive at the power and
bi-spectra we have presented in this work.


\subsection{Sensitivity of an interferometric array to HI power spectrum}

Let us assume that visibilities for a given baseline ${\bm u}$, $V_\nu({\bm u})$, 
corresponding to a statistically homogeneous process such as the HI signal or 
detector noise, are  measured for $n_\nu$ discrete frequency channels.
The measured visibilities can be Fourier transformed from frequency space to 
the delay space (see e.g. Ref.~\cite{2010ARA&A..48..127M}, and references 
therein) and correlated to obtain $\langle V_\tau({\bm u})\, V_\tau({\bm u})\rangle$,
where $\tau$ is the conjugate variable of the frequency difference between two maps. 
In discrete space, $\tau = n/\Delta\nu$, where $n$ runs between zero and $n_\nu/2$,
while $\Delta\nu = \delta\nu\, n_\nu$ is the total bandwidth and $\delta\nu$ is the 
channel width.  
The correlation in delay space can be related to the power spectrum as follows
(for a detailed derivation see, for instance, Ref.~\cite{Paul:2016blh}, and 
references therein)
\begin{equation}
\langle V_\tau({\bm u})\, V_\tau({\bm u})\rangle 
\simeq \f{\Delta\nu\, I_\nu^2\, \Omega_p}{r_\nu^2\, r_\nu'}\, P(k), 
\end{equation}
where $r_\nu$ is the coordinate distance, $r_\nu' \equiv \d r_\nu/\d\nu$, 
$\Omega_p$ is the primary beam,  and $I_\nu = k\, T_B/\lambda_0^2$. 
In delay space, different components of the  wave vector $k= \sqrt{{\bm k}_\perp^2 
+ k_\parallel^2}$, decomposed as modes on the plane of the sky~${\bm k}_\perp$, 
and along the line of sight~${\bm k}_\parallel$, are related to the parameters of
the radio interferometer as ${\bm k}_{\perp} \simeq  2\,\pi\, {\bf u}_\nu/r_\nu$
and $k_\parallel = 2\,\pi\, \tau/(\d r_\nu/\d\nu)$.

Assuming the visibility correlation to be dominated by the detector noise and a 
baseline coverage that yields uniform noise~\footnote{In an integration time 
$t_{\mathrm{int}}$, the total number of baselines is $N\,(N-1)\,t_{\mathrm{int}}/
(2\,t_{\mathrm{acc}})$, where $t_{\mathrm{acc}}$ is the time for one measurement. 
For high density of baselines, the instantaneous baseline coverage is complete
or it samples each UV pixel for a range of baselines (generally short baselines, 
e.g. MWA for baselines shorter than roughly  $200\,\lambda_0$~\cite{wayth2018phase}). 
Therefore, the total number of independent UV pixels remains $N\,(N-1)/2$ after 
$t= t_{\mathrm{int}}$. 
This yields uniform noise in each pixel in both UV and real space.}, one can obtain 
the error on the brightness temperature fluctuations  from  the detector noise (for
each primary beam and for non-redundant baselines):
\begin{equation}
\delta\l(\f{T_B^2\, k^3\, P(k)}{2\,\pi^2}\r) 
\simeq \f{T_{\mathrm{sys}}^2\, \Omega_p\, r_\nu'\, r_\nu^2\, \delta\nu\, k^3}{2\,\pi^2\,
f^{1/2}(k)\, N\,\Delta\nu\, t_{\mathrm{int}}},\label{eq:tbsens}
\end{equation}
where $T_{\mathrm{sys}}$ is the system temperature, $N$ is the number of antennas, 
and $f(k)$ is the fraction of baseline in the range~$k$ and $(k+\d k)$. 
Current (e.g. LOFAR) and future low-frequency radio interferometers have the capability 
to form and image in multi-beams. 
Multiple primary beams help to improve both the detector sensitivity and the sample
variance. 
For $n_p$ primary beams, the brightness temperature variance derived in Eq.~(\ref{eq:tbsens}) 
decreases by a factor of~$n_p^{1/2}$. 

It is our aim that the interferometer be sensitive to wavenumbers $k \simeq 1$--$10\, 
\mathrm{Mpc}^{-1}$. 
We consider two cases:~(a)~$k_\perp \gg k_\parallel$, and (b)~$k_\parallel \gg k_\perp$.
When $k_\perp \gg k_\parallel$, the baseline is $u \simeq k\, r_\nu/(2\,\pi)$. 
For $z\simeq 50$, this gives a baseline length ($b = u\, \lambda_0$) of nearly 
$200\,\mathrm{km}$ for $k\simeq 10 \, \mathrm{Mpc}^{-1}$. 
In three dimensions, $f(k) \propto k^3$, which causes the brightness temperature 
sensitivity to scale as $k^{3/2}$ (see e.g. Ref.~\cite{Furlanetto:2006wp}). 
However, it is not easy to achieve uniform coverage in this case as the number of
telescopes needed to achieve it scales as $k^2$. 
For instance, if uniform UV coverage could be achieved with 100 telescopes for $k 
= 0.1\, \mathrm{Mpc}^{-1}$ (e.g. MWA configuration, see Ref.~\cite{wayth2018phase}), 
then nearly $10^4$ telescopes would be needed to reach the same goal for $k = 1\, 
\mathrm{Mpc}^{-1}$. 
We therefore consider another novel method which is more suitable for the detection 
of the HI signal at small scales, viz. when $k_\parallel \gg k_\perp$. 
In this case, $k \simeq k_\parallel$ and the baseline length does not play a role in 
determining the sensitivity of the array to a given Fourier mode. 
In this case, $f_k = 1$ as  all baselines contribute equally to the sensitivity for
any scale for which $k_\parallel \gg k_\perp$.  

We note that better sensitivity might be achieved in case (b) as $f(k) =1$. 
However, the sensitivity scales as $k^3$ for case~(b)  whereas it scales as $k^{3/2}$ 
for case~(a), so case~(a) might be the method of choice at large  scales.
We choose the following parameters for computing the sensitivity at $z= 50$ for case~(b): integration time, $t_{\mathrm{int}} = 3\times 10^7$; channel width,  $\delta \nu = 10 \, 
\mathrm{kHz}$; total bandwidth, $\Delta\nu = 10 \, \mathrm{MHz}$; number of antennas,  
$N= 5000$; system temperature, $T_{\mathrm{sys}} = 10^4\, \mathrm{MHz}$;  primary beam, 
$\Omega_p = (4/57)^2$; number of beams, $n_p = 100$.  
This yields a brightness temperature variance $\delta(T_B^2\, k^3\, P(k)/(2\,\pi^2)) 
\simeq (1\, \mathrm{mK})^2$ at $k = 3\, \mathrm{Mpc}^{-1}$. 
These parameters are  within the reach of current technologies and might be realizable  
by SKA-II. 
However, ground-based telescopes might not work well at $\nu = 30\, \mathrm{MHz}$ needed 
to detect the signal due to ionospheric effects. 
Future space-based and lunar radio interferometers will evade this 
constraint~\cite{Furlanetto:2019jso,Cole:2019zhu,Koopmans:2019wbn}. 
We note that the sensitivity might be improved further with redundant baselines: 
if each baseline has redundancy $M$, the  total number of independent baselines 
is nearly $N^2/(2\,M)$. 
While the non-redundant baselines yield incoherent addition, the redundant baselines 
contribute coherently. 
This results in sensitivity on the noise power spectrum being better by a factor~$M^{-1}$. 

\bibliographystyle{apsrev4-2}
\bibliography{references.bib}